\documentclass[10pt,twocolumn]{IEEEtran} 

\usepackage{amsmath,url}
\usepackage{epsfig,amssymb,amsbsy,verbatim,array,enumerate}
\usepackage{pstricks,psfrag,theorem,cite,paralist,bm,subfigure}

\let\intern=\iftrue

\def\argmax{\operatorname{arg~max}}
\newcommand{\argmin}{\operatornamewithlimits{arg\ min}}
\def\figref#1{Fig.\,\ref{#1}}%
\def\E{\mathbb{E}}
\def\P{\mathbb{P}}
\def\R{\mathbb{R}}

\def\N{\mathbb{N}}

\def\ie{{\em i.e.}}

\def\var{\operatorname{var}}

\def\sir{\mathsf{SIR}}
\def\sfn{\mathsf{SFN}}

\def\sinc{\operatorname{sinc}}
\def\dd{\mathrm{d}}

\def\sf{\mathsf{SF}}
\def\misr{\mathsf{MISR}}
\def\misf{\mathsf{MISF}}
\def\isr{\mathsf{ISR}}
\def\one{\mathbf{1}}

\def\mh{{\rm{MH}}}

\newtheorem{lemma}{Lemma}

\newtheorem{definition}{Definition}

\newtheorem{proposition}{Proposition}
\newtheorem{conjecture}{Conjecture}

\newlength{\figwidth}


\begin{document}
\title{SIR Analysis via Signal Fractions} 
\author{Martin Haenggi\\
Dept.~of Electrical Engineering\\University of Notre Dame}
\maketitle
\begin{abstract}
The analysis of signal-to-interference ratios (SIRs) in wireless networks is instrumental to derive important
performance metrics, including reliability, throughput, and delay. While a host of results on SIR distributions
are now available, they are often not straightforwards to interpret, bound, visualize, and compare.
In this letter, we offer an alternative path towards the analysis and visualization of the SIR distribution.
The quantity at the core of this approach is the {\em signal fraction} (SF), which is the ratio of the signal power to the total received power.
A key advantage is that the SF is constrained to $[0,1]$. We exemplify the benefits of the SF-based approach
by reviewing known results for Poisson cellular networks. In the process,  we derive new approximation and bounding techniques
that are generally applicable.
\end{abstract}
\section{Introduction}
The signal-to-interference ratio (SIR) at a receiver is defined as $\sir\triangleq S/I$, where $S$ is the signal power
(emitted by the desired transmitter), and $I$ is the total interference power (emitted by all other concurrent transmitters).
Its distribution is an important performance metric in wireless networks, characterizing the reliability of a transmission
in an interference-limited network. This letter shows that it is often advantageous to focus on {\em signal fractions} instead of
SIRs, for both analysis and visualization. 
\subsection{Definition}
\begin{definition}[Signal fraction]
The {\em signal fraction} (SF) is defined as the ratio of the signal power to the total
received power, \ie,

\[ \sf\triangleq \frac{S}{S+I}. \]
\end{definition}
Hence, defining $T(x)\triangleq x/(1+x)$, we have $\sf=T(\sir)$ and $\sir=T^{-1}({\sf})$, \ie,
\[ \sf=\frac\sir{1+\sir};\quad \sir=\frac\sf{1-\sf} . \]

$T$ is a homeomorphism between $\R^+=\{x\in\R\colon x\geq 0\}$ and $[0,1)$, with fixed point $0$.

Letting $F_X$ denote the cumulative distribution function (cdf) of the random variable $X$,
$\bar F_X$ its complement (ccdf),
and $f_X$ the corresponding probability density function (pdf),
we have the relationships
\[ \bar F_{\sir}(\theta)=\bar F_{\sf}(T(\theta)); \quad \bar F_{\sf}(t)=\bar F_{\sir}(T^{-1}(t)) .\]

For the pdfs, $f_{\sir}(\theta)=f_{\sf}(T(\theta))\frac{\dd T(\theta)}{\dd\theta}$, hence
\[ f_{\sir}(\theta)=\frac{f_{\sf}(\theta/(1+\theta))}{(1+\theta)^2} ;\quad f_{\sf}(t)=\frac{f_{\sir}(t/(1-t))}{(1-t)^2}. \]

\subsection{Visualization and MH Units}
Since the support of the SIR is $\R^+$, its distribution cannot be fully shown on a linear scale. Switching to 
a logarithmic scale helps somewhat as it compresses high SIR values, but now the support is the entire $\R$.
In contrast, the SF is supported on $[0,1]$, which makes it easy to plot in full.
Based on the map $T$, we define a new unit, called the {\em M\"obius\footnote{$T$ is also a (parabolic) M\"obius transformation.}
homeomorphic} unit, abbreviated to $\mh$.
For $x\in [0,1)$,
$x~\mh = \frac x{1-x}$.
For comparison, the dB unit is defined as $x~{\rm dB}=10^{x/10}$.

\begin{figure}
{\epsfig{file=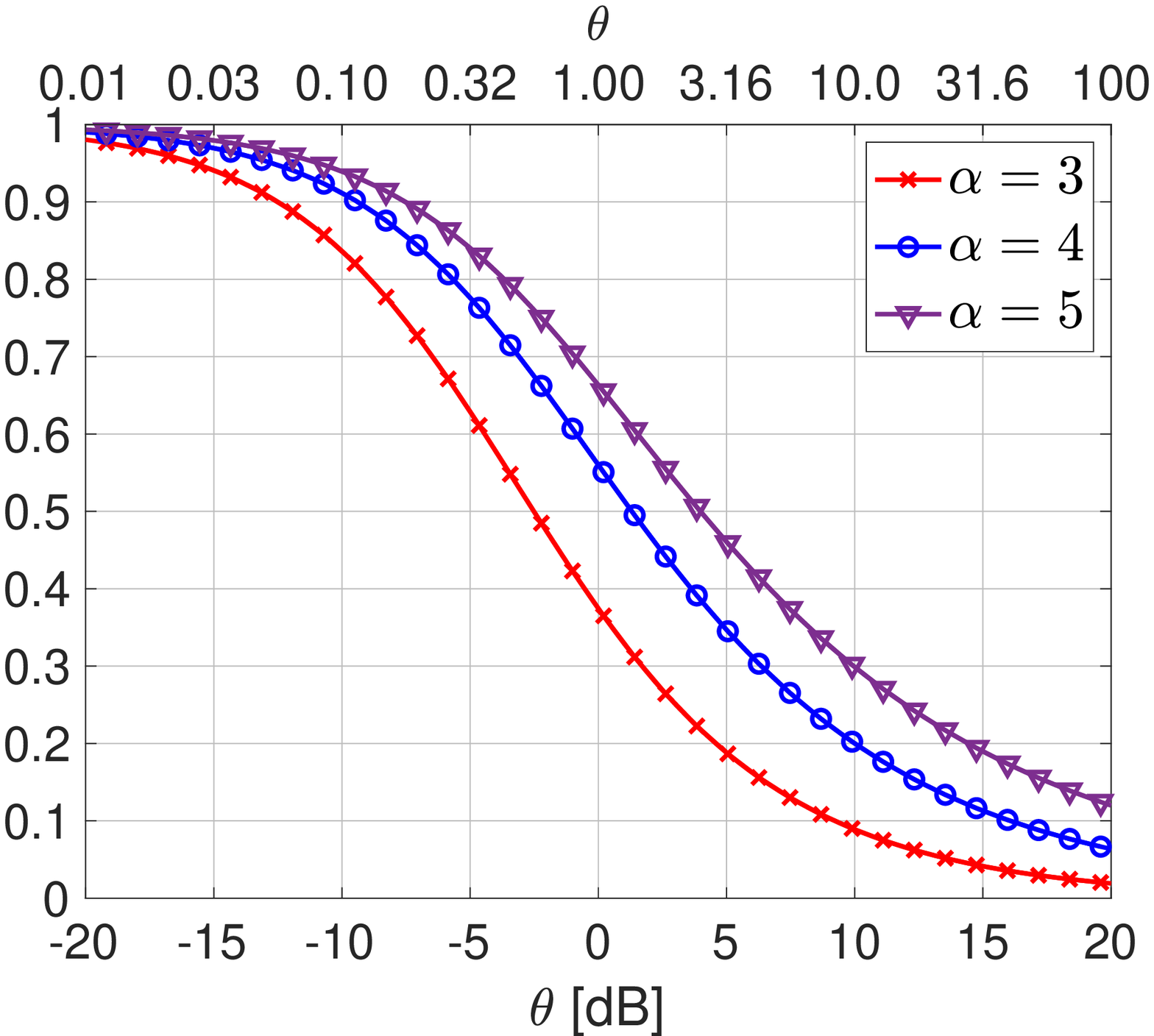,width=.475\columnwidth}}\hfill
{\epsfig{file=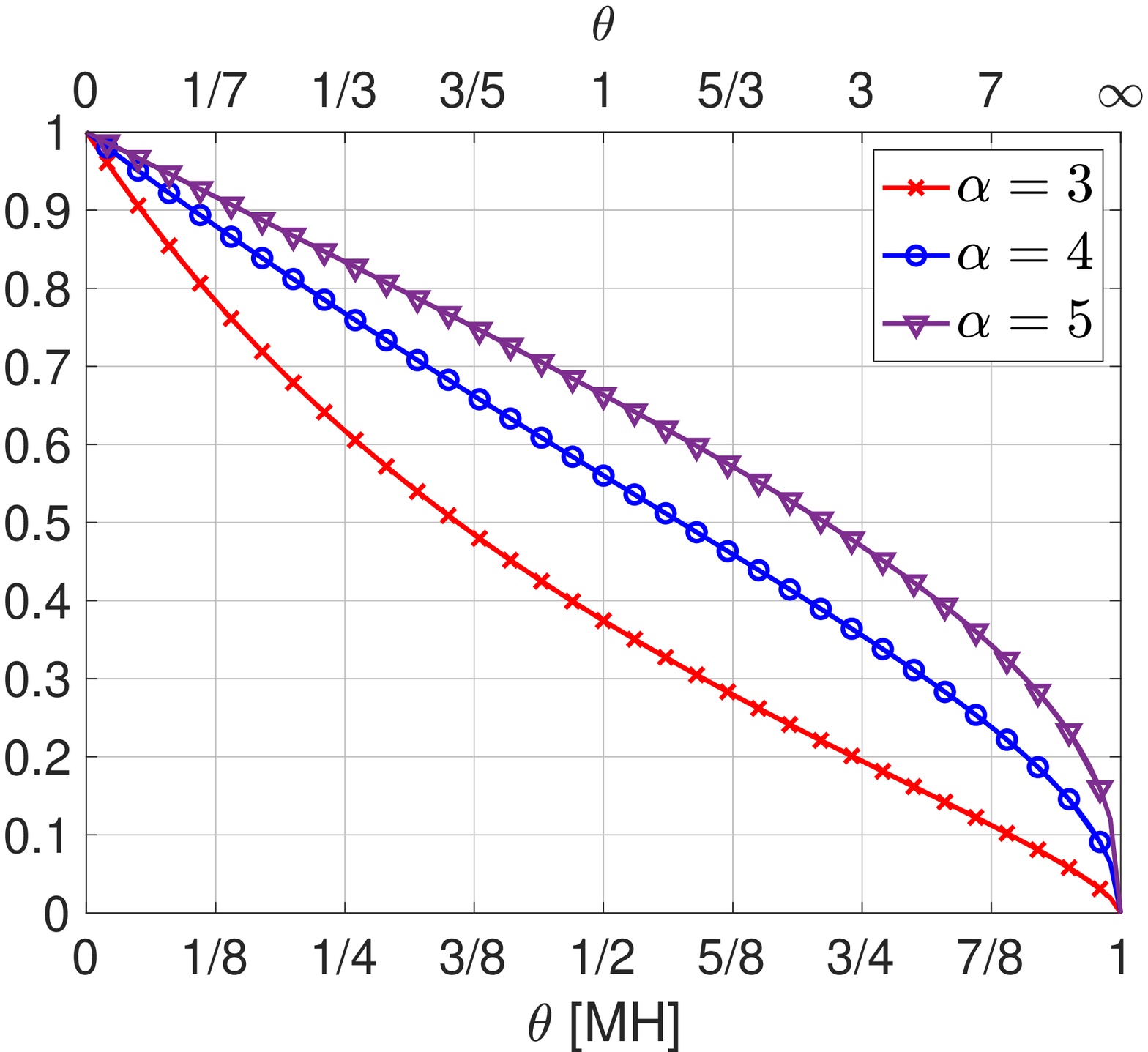,width=.475\columnwidth}}
\caption{SIR distribution $\bar F_{\sir}(\theta)$ for Poisson networks (per \eqref{ps} and \eqref{ccdf}) with different path loss exponents
in units of dB (left) and MH (right). The top axis gives the
corresponding values of $\theta$ in standard (linear) units.}
\label{fig:units}
\end{figure}

Thus equipped, we can write $\theta=T(\theta)~\mh$. \figref{fig:units} shows SIR ccdfs for Poisson cellular networks (see Sec.~\ref{sec:fad})
 in units of dB and MH.

An advantage of the MH scale vs.~the dB scale is that $\theta\sim T(\theta)$, $\theta\to 0$, \ie, $\theta \sim \theta~\mh$.
Hence, in the important high-reliability regime, $T$ is linear, which means that the ccdf directly
reveals the tradeoff between rate and reliability. The (normalized) rate (in nats/s/Hz) is given by
$\log(1+\theta)\sim\theta$ or $-\log(1-t)\sim t\sim \theta$.
Put differently, the MH unit has higher discriminative power for high reliabilities than the dB unit.

\subsection{Poisson Cellular Network Model}
In the following two sections, we focus on the downlink in Poisson cellular networks. We let $\Phi\subset\R^2$ be a
stationary Poisson point process (PPP) of arbitrary positive intensity and focus on the typical user located at
the origin. All our results also hold for the homogeneous independent Poisson (HIP) model, consisting
of the union of an arbitrary number of PPPs of arbitrary densities where the base stations of each tier
transmit at the same arbitrary power levels.

If $y\in\Phi$ is the desired transmitter, the signal fraction is
\begin{equation}
 \sf_y = \frac{h_y\ell(y)}{\sum_{x\in\Phi} h_x \ell(x)} .
\label{sfy}
\end{equation}
We let $\ell(x)=\|x\|^{-\alpha}$, where $\alpha=2/\delta$ is the path loss exponent. $(h_x)_{x\in\Phi}$ are 
independent and identically distributed (iid) random variables with $\E(h_x)=1$ representing fading. 

We will study two cases: In Section \ref{sec:fad}, we focus on networks with fading and nearest-base station association, \ie,
$\sf=\sf_y$ where $y=\argmin\{x\in\Phi\colon \|x\|\}$. We denote this case by NBA-$m$, where $m$ is the Nakagami-$m$
fading parameter.
Section \ref{sec:opp} addresses the no-fading case, or, equivalently, the case of
instantaneously-strongest base station association (ISBA) with arbitrary fading\footnote{For ISBA, it is known that
the SIR distribution does not depend on the fading statistics \cite{net:Zhang14tit}, and without fading, ISBA and NBA
are identical.}.
In this case, $\sf=\sf_y$ where $y=\argmax\{x\in\Phi\colon h_x\ell(x) \}$ or, equivalently, setting all $h_x=1$ and selecting
$y=\argmin\{x\in\Phi\colon \|x\|\}$.

\section{Signal Fraction with Fading and Nearest-Base Station Association}
\label{sec:fad}
We first focus on Rayleigh fading where the $h_x$ are exponential, \ie, NBA-$1$.
\subsection{Exact Distribution}
The SIR distribution is \cite{net:Zhang14twc} 
\begin{equation}
  \bar F_{\sir}(\theta)=\frac1{\,_2F_1(1,-\delta;1-\delta,-\theta)},
  \label{ps}
\end{equation}
where $\,_2F_1$ is the Gauss hypergeometric function.
It follows that
the cdf of the $\sf$ is given by $\bar F_{\sf}(t)=\bar F_{\sir}(t/(1-t))$, which can be expressed more compactly as
\begin{equation}
  \bar F_{\sf}(t)=
  \frac{1}{(1-t)\,_2F_1(1,1;1-\delta,t)}. 
 \label{ccdf}
 \end{equation}
 This expression, compared with \eqref{ps}, has the advantage that the last argument of the hypergeometric
function does not exceed $1$, which speeds up the evaluation.

\subsection{Asymptotics and Approximations}
\subsubsection{Rational Approximation}
The ccdf of the SF in \eqref{ccdf} can be expressed as
\[ \bar F_{\sf}(t)=\frac{\sum_{n=0}^\infty t^n}{\sum_{n=0}^\infty a_n t^n} ,\]
where $a_n=\Gamma(n+1)\Gamma(1-\delta)/\Gamma(n+1-\delta)$. Truncations of the infinite series to
numerator and denominator polynomials of order $s$ yield simple rational (Pad\'e-type) approximations
whose
first $s$ derivatives at $t=0$ match those of the exact expression, \ie, they are all asymptotically exact
as $t\to 0$.
For example, for $s=2$,
\[ \bar F_{\sf}(t)\sim \frac{1+t+t^2}{1+t/(1-\delta)+2t^2/((1-\delta)(2-\delta))},\quad t\to 0.\]

\subsubsection{Polynomial Approximation}
The slope of the cdf at $0$ is $f_{\sf}(0)=\misr$, consistent with the known result
$\P(\sir\leq\theta)\sim\misr\,\theta$, $\theta\to 0$ \cite{net:Haenggi14wcl}.
As a result, $\bar F_{\sf}(t)\sim 1-\misr\,t$ is a good approximation for reliabilities of $0.8$ and above
(\ie, $t\leq 0.2/\misr$).

Adding the second-order term, we obtain
\begin{equation}
   \bar F_{\sf}(t)\sim 1-\misr\,t+\left(\frac{\misr^2-\delta}{2-\delta}\right)t^2,\quad t\to 0.
   \label{ccdf_2nd}
 \end{equation}
If $\misr^2-\delta>0$, 
the ccdf is locally convex at $t=0$. This holds if $\delta>(3-\sqrt{5})/2\approx 0.382$ (equivalently, if
$\alpha<4/(3-\sqrt 5)\approx 5.24$), and it implies that $1-\misr\,t$ is a lower bound while \eqref{ccdf_2nd}
is an upper bound. Conversely, for $\delta<0.382$ ($\alpha>5.24$), both first- and
second-order asymptotics are upper bounds.
We can conclude that in most practical cases, $1-\misr\,t$ is a lower bound.

Applied to the SIR, we immediately have $F_{\sir}(\theta)\sim \misr\,\theta/(1+\theta)$, which is a significantly
better approximation than just $\misr\,\theta$. Generally, $T$ turns polynomials
for the SF into rational functions for the SIR of the same order, with improved accuracy.
In comparison, the Pad\'e approximation in \cite{net:Nagamatsu14spaswin} requires the calculation of twice
as many derivatives as the approach via the SF.

\begin{figure}
\subfigure[SF, $\alpha=3$.]{\epsfig{file=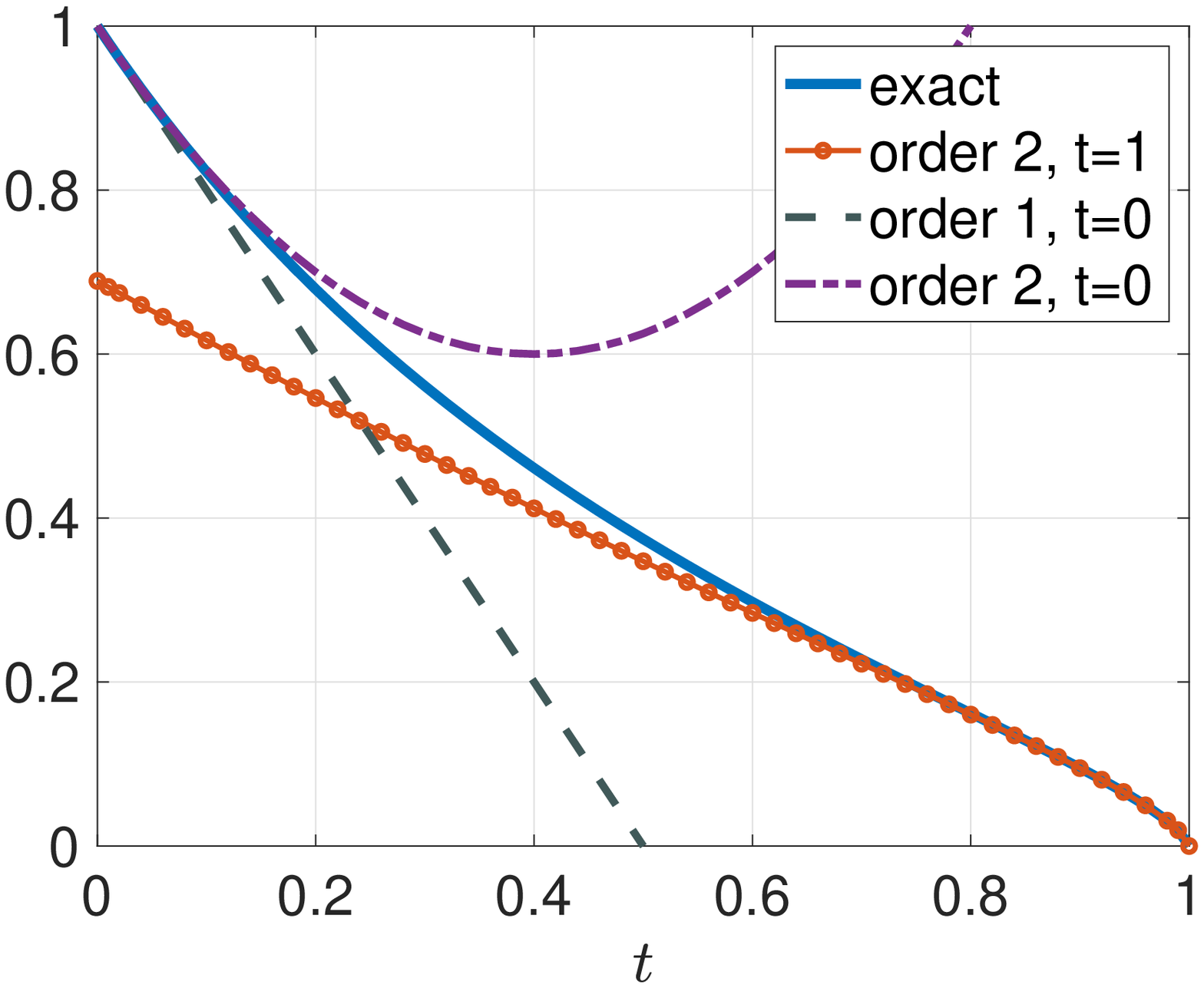,width=.48\columnwidth}}\hfill
\subfigure[SIR, $\alpha=3$.]{\epsfig{file=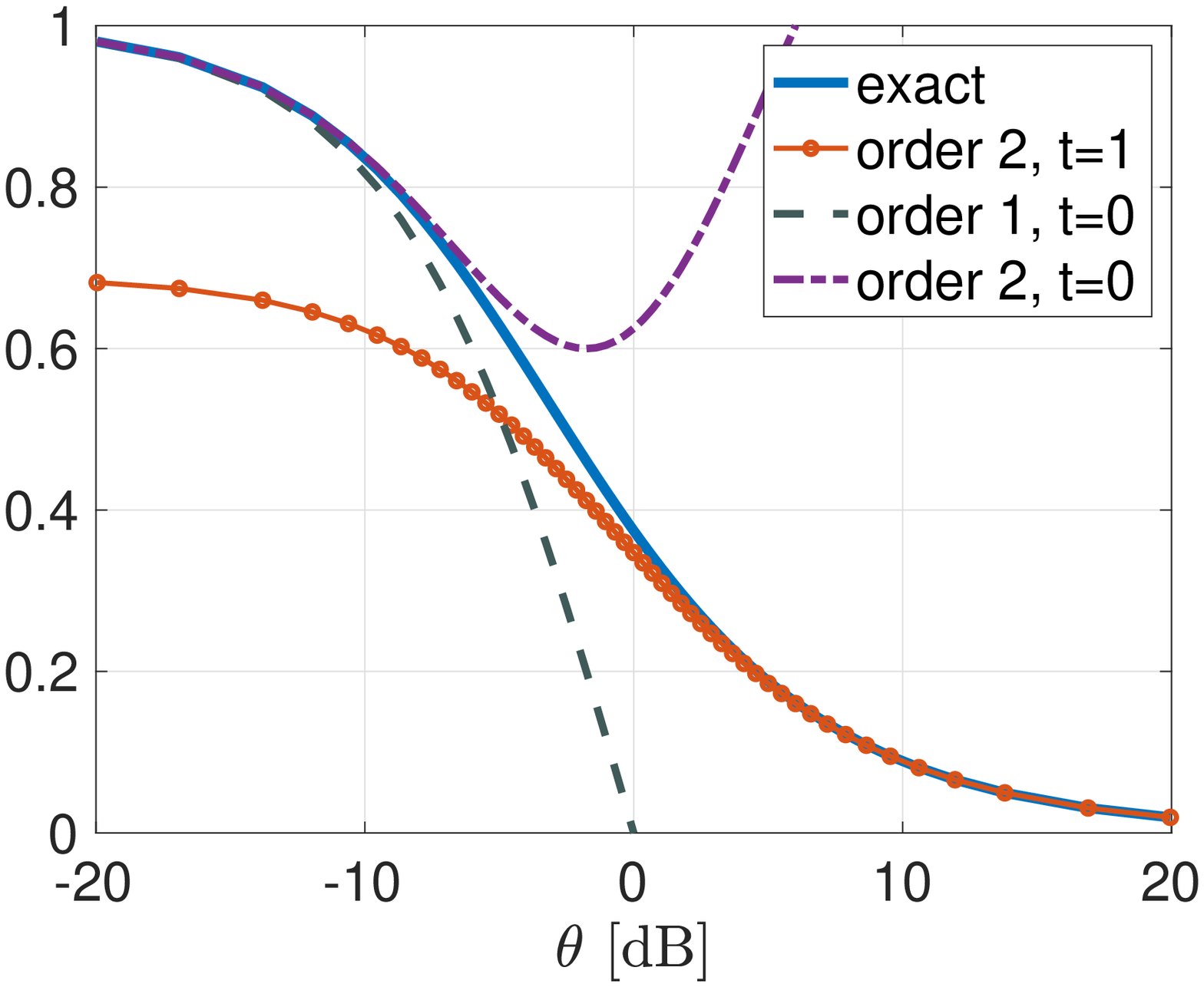,width=.48\columnwidth}}
\subfigure[SF, $\alpha=4$.]{\epsfig{file=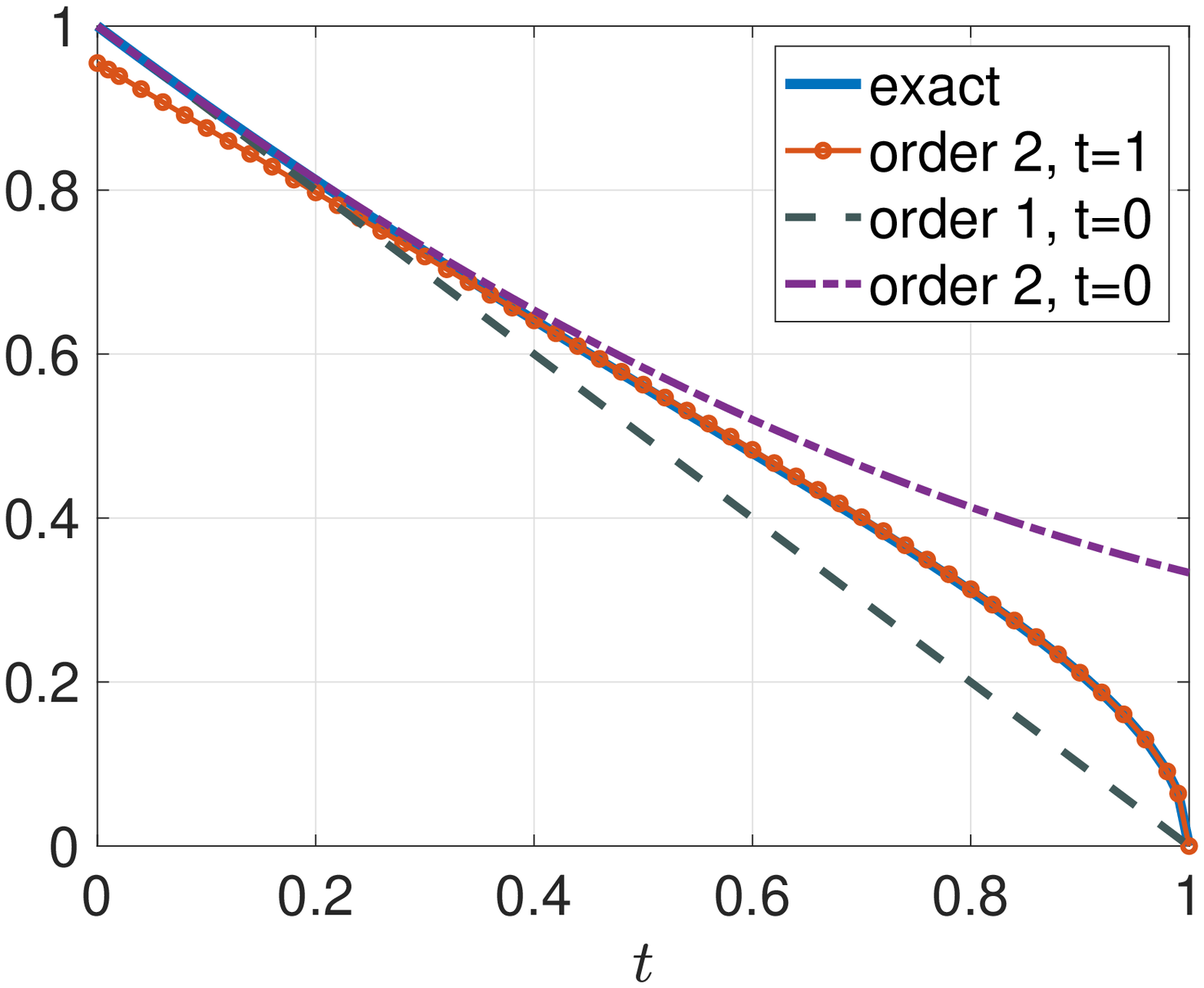,width=.48\columnwidth}}\hfill
\subfigure[SIR, $\alpha=4$.]{\epsfig{file=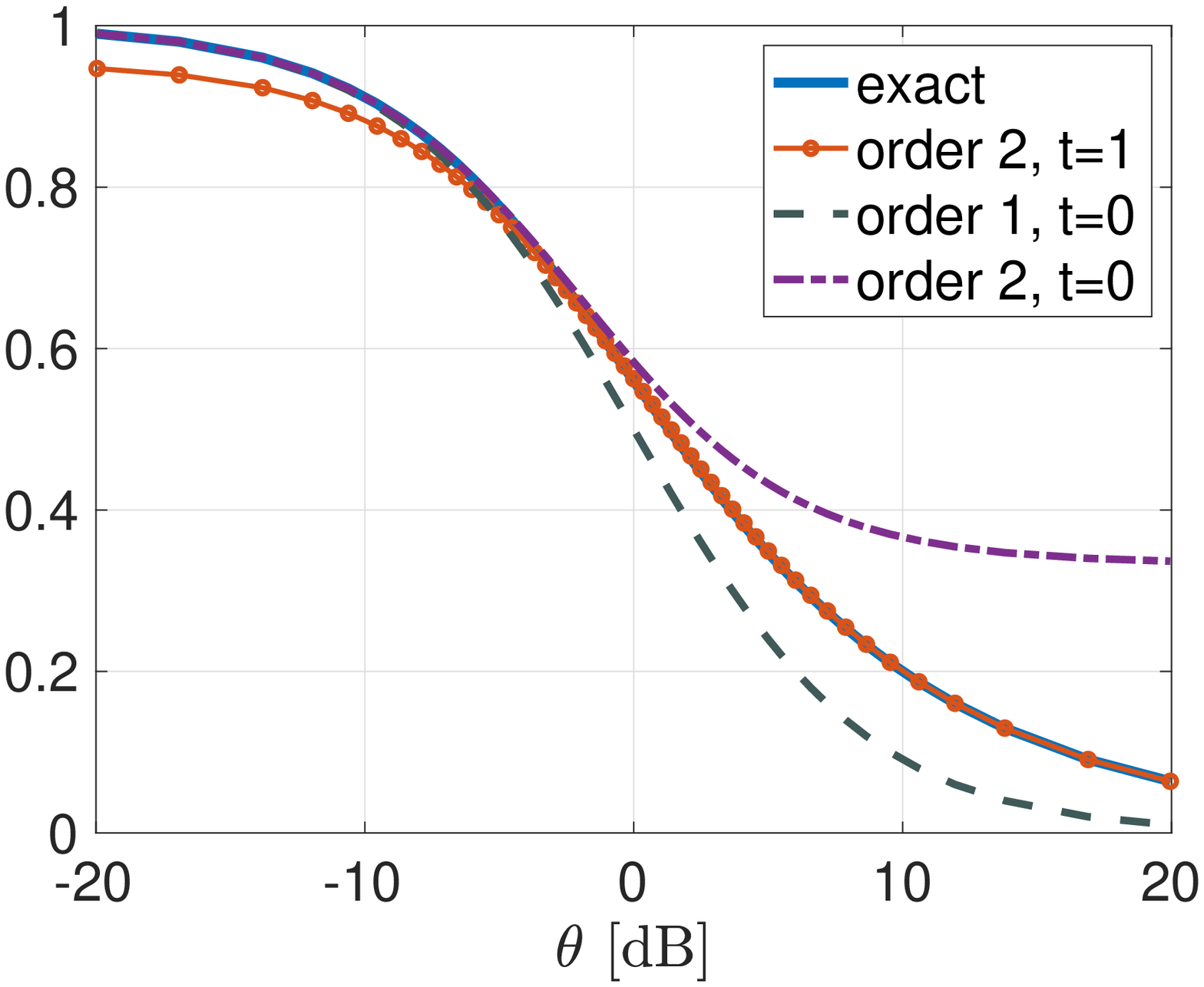,width=.48\columnwidth}}
\caption{SF and SIR ccdfs and approximations for $\alpha=3,4$. The first-order approximation at $t=0$ is
$1-\misr\, t$ and the second-order one is given in \eqref{ccdf_2nd}. The second-order approximation at $t=1$ is given in 
\eqref{asymp12}.}
\label{fig:approx4}
\end{figure}

\figref{fig:approx4} illustrates the exact results and different approximations for the ccdfs of the SF and their
application to the ccdfs of the SIR.

\subsubsection{Series Expansion at $t=1$}
From \eqref{ccdf} we can derive the second-order series expansion
\begin{equation}
 \bar F_{\sf}(t)\sim \sinc(\delta)(1-t)^\delta(1+\delta(1-t)) ,\quad t\to 1,
 \label{asymp12}
\end{equation}
where $\sinc(x)\triangleq \sin(\pi x)/(\pi x)$.
It turns out to be a very good approximation for at least $t>2/3$, see \figref{fig:approx4}. Removing the factor $1+\delta(1-t)$,
 the first-order expansion is obtained.
This asymptotic result shows that the slope at $t=1$ is always infinite, \ie, $f_{\sf}(1)=\infty$.

\subsubsection{Beta Approximation}
For the SIR, there is no simple distribution that closely resembles the entire actual distribution.
For the SF, the beta distribution with pdf $f_\beta(t)=t^{p-1}(1-t)^{q-1}/{\rm B}(p,q)$, where
${\rm B}$ is the beta function, is a natural candidate.
However, merely requiring $f_{\sf}(0)=\misr$ fixes both parameters, namely
$p=1$ and $q=\misr$, and the resulting $f_{\beta}(t)=\misr(1-t)^{\misr-1}$ does not match the asymptotics
at $t=1$. For instance, if $\misr>1$, $f_\beta(1)=0$ instead of $\infty$, and for $\misr=1$, it is just the
uniform distribution.

To have more degrees of freedom, we turn to the five-parameter generalized beta distribution put forth in \cite{net:McDonald95}.
With a support of $[0,1]$ and $0<f_{\sf}(0)<\infty$,
one of the parameters 
can be eliminated, resulting in the four-parameter pdf
\[ f_{{\rm GB}}(t; a,b,p,q)\triangleq \frac{a(1-t^a)^{q-1}}{b{\rm B}(p,q) (1+(b^{-a}-1)t^a)^{p+q}}, \]
with $a=1/p$.
Since $f_{\sf}(0)=\misr$, we have $b=(\misr\, p\, {\rm B}(p,q))^{-1}$, which leaves
the two parameters $p$ and $q$ to match other statistics.

A simple option is to match the $\Theta((1-t)^{\delta-1})$ asymptotics at $t=1$. It yields
$a=p=1$ and $q=\delta$, and thus $b=1-\delta$, resulting in 
\begin{equation}
  \tilde f_{\rm GB}(t)=\frac{\mu}{(1-t)^{1-\delta}(1+\mu t)^{1+\delta}} ,
 \label{betaA}
 \end{equation}
 where $\mu=\misr$. It satisfies $\tilde f_{\rm GB}(t)\sim\delta(1-\delta)^\delta (1-t)^{\delta-1}$, $t\to 1$, which is slightly
 larger\footnote{The maximum gap between the pre-constants is $0.045$ at $\delta=0.65$.} than the actual $\delta\sinc\delta(1-t)^{\delta-1}$ from \eqref{asymp12}. We call the resulting approximation of the SF and SIR distributions the {\em beta-based simple tight} (BEST) approximation.
 It is formally stated in terms of the ccdfs in the following proposition.
 \begin{proposition}[BEST approximation]
 For Rayleigh fading, the SF and SIR distributions are tightly approximated by
 \begin{equation}
 \bar F_{\sf}^{\small \text{\rm BEST}}(t)=\left(\frac{1-t}{1+\mu t}\right)^\delta;\; \bar F_{\sir}^{\small\text{\rm BEST}}(\theta)=\big(1+(1+\mu)\theta\big)^{-\delta} ,
 \label{best}
 \end{equation}
 respectively, where $\mu=\misr=\delta/(1-\delta)$.
 \end{proposition}

\begin{figure}
\centerline{\subfigure[SIR ccdf in dB]{\epsfig{file=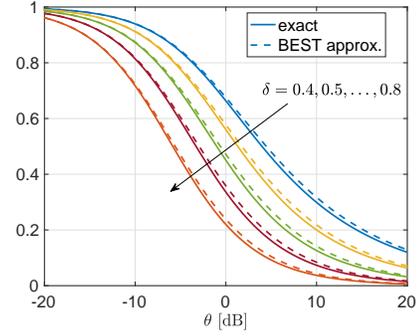,width=.75\figwidth}}}
\centerline{\subfigure[SIR ccdf in MH (or SF ccdf)]{\epsfig{file=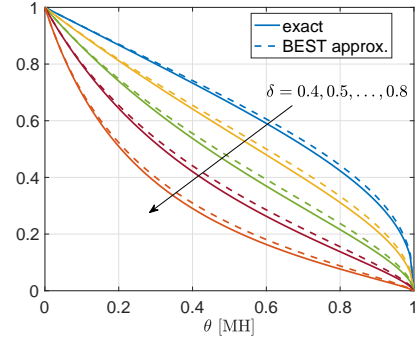,width=.75\figwidth}}}
\caption{SIR ccdfs and BEST approximations \eqref{best} for $\delta=0.4,\ldots,0.8$, corresponding to a range of $\alpha$ from $5$ to $2.5$.}
\label{fig:best}
\end{figure}
\figref{fig:best} shows the exact SIR ccdfs and the BEST approximations for a range of $\delta$ values. 
The accuracy of the very simple approximation is remarkable. 
Its inverse is equally simple, which makes it easy to find the SF or SIR thresholds
for a given target reliability.

With a bit more effort we can determine $p$ and $q$ by matching the first and second moments $M_1$ and $M_2$, given by
 \cite[Eqn.~(2.10)]{net:McDonald95}
\[ M_k=\frac{b^k{\rm B}((k+1)p,q)}{{\rm B}(p,q)} \,_2F_1((k+1)p, kp; (k+1)p+q; 1-b^a). \]
This way, we obtain
\begin{equation}
\hat f_{\rm GB}(t)=f_{\rm GB}\big(t; 1/p,(\mu p{\rm B}(p,q))^{-1},p,q\big)
\label{beta5}
\end{equation}
with $p$ and $q$ chosen such that $M_1=\E(\sf)$ and $M_2=\E(\sf^2)$.
Table \ref{table:beta} shows the numerically obtained values of $b$, $p$, and $q$ for 
$\alpha=3,4,5$. The resulting approximations are virtually indistinguishable from the exact distributions.

\begin{table}
\[ \begin{array}{|c|c|c|c|}
\hline
\delta & b & p & q \\\hline
2/5 & 0.7160 &0.7385  &  0.4164 \\
1/2 &0.5554 & 0.8648 &0.5276 \\
2/3 & 0.3598 & 0.9296  & 0.7089 \\\hline
\end{array} \]
\caption{Values of $b$, $p$ and $q$ for different $\delta$ for the generalized beta approximation in \eqref{beta5}.}
\label{table:beta}
\end{table}

\begin{figure}
\centerline{\epsfig{file=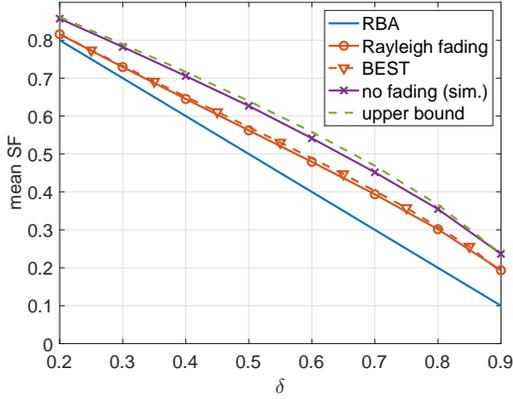,width=.95\figwidth}}
\caption{Mean signal fraction for random base station association (RBA) \eqref{random}, Rayleigh fading \eqref{ccdf}, the BEST approximation \eqref{best}, no fading (simulated), and the upper bound \eqref{sf_upper}. The range of $\delta$ corresponds to $\alpha\in [2.25,10]$.} 
\label{fig:mean_sf}
\end{figure}

 \figref{fig:mean_sf} shows the mean signal fractions for Rayleigh fading, obtained from \eqref{ccdf}, the BEST
 approximation \eqref{best}, a simulation result for the no-fading case (ISBA---see Sec.~\ref{sec:opp}), an upper bound for it, and the random base station association
 scheme discussed in Subsec.~\ref{sec:random}.

\subsection{Other Fading Models}
\label{sec:fading}
For NBA-$m$, $F_h(x)\sim c_m x^m$, $x\to 0$, where $c_m=m^{m-1}/\Gamma(m)$.
As shown in \cite{net:Ganti16twc},
\[ \bar F_{\sir}(\theta)\sim 1-c_m\theta^m \E(\isr^m),\quad\theta\to 0 ,\]
where $\isr=I/\E_h(S)$ is the interference-to-(average) signal ratio (ISR), \ie, $\E\,\isr=\misr$.
The $m$-th moment for $m\in\N$ of the ISR for arbitrary fading is given in \cite[Thm.~2]{net:Ganti16twc}.
For $m=2$, for example,
$\E(\isr^2)=2\,\misr^2+\frac{\delta\E(h^2)}{2-\delta}$. This means that for NBA-$2$, where $c_2=2$
and $\E(h^2)=3/2$,
\[ \bar F_{\sf}(t)\sim 1-\frac{\delta(3+2\delta-\delta^2)}{(1-\delta)^2(2-\delta)} t^2 ,\quad t\to 0.\]
Conversely, the asymptotics as $t\to 1$ do not depend on the fading model, \ie,
the tail for NBA-$m$ is $\sinc(\delta)(1-t)^\delta$ (see \eqref{asymp12}) for any $m>0$ \cite[Lemma 6]{net:Ganti16twc}.
Consequently, a beta approximation similar to \eqref{betaA} but with $p=m$ is expected to perform well.

\section{Signal Fractions Without Fading}
\label{sec:opp}
\subsection{The Path Loss Point Process}
For a PPP $\Phi\subset\R^2$ of intensity $\lambda$, let the {\em path loss point process} (PLP) be defined as
$\Xi\triangleq\{x\in\Phi\colon \|x\|^\alpha/V_x\}\subset \R^+$, where the $V_x$ are iid with
$\E(V^\delta)<\infty$, representing shadowing and/or fading. 
The PLP is itself Poisson and has the intensity measure $\Lambda([0,r])=\lambda \pi \E(V^\delta) r^\delta$
\cite{net:Zhang14tit}.
Scaling the density does not affect the SF or SIR distributions, so we can equivalently work with a PLP of intensity measure
$\Lambda([0,r])=r^\delta$, ignoring any shadowing or fading\footnote{As pointed out earlier,
ISBA performs exactly like NBA-$\infty$.}.

If the elements of $\Xi=\{\xi_1,\xi_2,\ldots\}$ are ordered (increasingly), their pdfs are \cite[Lemma 3]{net:Zhang14tit}
\[ f_{\xi_k}(x)=\frac{\delta x^{k\delta-1}}{\Gamma(k)}e^{-x^\delta} .\]

In  \cite{net:Keeler14wcl},
the signal-to-total-received-power ratio process is introduced 
and shown to be a Poisson-Dirichlet process with parameters $(\delta,0)$. 
It is defined as
$\Psi\triangleq \{\xi\in\Xi\colon\xi^{-1}/P\}\subset [0,1]$, where $P=\sum_{\xi\in\Xi} \xi^{-1}$
is the total received power. The elements of $\Psi=\{\sf_{k}\}_{k\in\N}$, when ordered decreasingly, are the signal fractions
when the user is served by the $k$-th strongest base station.

\subsection{Distribution of Signal Fractions}
We first present a lemma summarizing some results on the statistics of the signal fractions.
\begin{lemma}
For $i\in\N$,
\begin{equation}
 \E\left(\frac{\sf_{i}}{\sf_{1}}\right)=\frac{\Gamma(i)\Gamma(1+1/\delta)}{\Gamma(i+1/\delta)}, 
 \label{sfirat}
 \end{equation}
and
\begin{equation}
   \E\log(\sf_{i+1})=\E\log(\sf_1)-H_i/\delta ,
   \label{harm1}
\end{equation}
where $H_i=1+2^{-1}+\ldots+i^{-1}$ is the $i$-th harmonic number.
Moreover,
letting 
\begin{equation}
g_n(t)\triangleq \frac{(t^{-1}-1)^{n\delta}}{\Gamma(1+n\delta)(\Gamma(1-\delta))^n}
\label{gn}
\end{equation}
and
$\sf_{\Sigma}^{(n)}\triangleq\sum_{i=1}^n \sf_i$, we have
\begin{equation}
 \P(\sf_n+t\sf_{\Sigma}^{(n-1)}>t)=g_n(t),\quad\tfrac12\leq t\leq 1 .
\label{sic}
\end{equation}

\end{lemma}
\begin{IEEEproof}
As shown in \cite{net:Keeler14wcl}
the ratio $R_i\triangleq \sf_{i+1}/\sf_{i}$ has the cdf
\[ \P(R_i\leq r)=r^{i\delta},\quad 0\leq r\leq 1, \]
and the $R_i$ are independent with $\E(R_i)=i\delta/(1+i\delta)$. 
\eqref{sfirat} then follows from $\sf_{i}/\sf_1=\prod_{k=1}^{i-1} R_i$.
Similarly, \eqref{harm1} follows from $\E(\log R_i)=-1/(\delta i)$ and summation.

Lastly, \eqref{sic} is obtained by rewriting $\P(\xi_n^{-1}/\sum_{k=n+1}^\infty \xi_k^{-1} \geq \theta)$, $\theta\geq 1$, given in  \cite[Thm.~1]{net:Zhang14tit}, in terms of signal fractions.

\end{IEEEproof}
{\em Remarks.}
\begin{itemize}
\item \eqref{sfirat} also obtained by integrating the ccdf of $\sf_{i}/\sf_1$ given in \cite[Lemma 3]{net:Zhang14twc}.
\item The expectations in \eqref{sfirat} add up to $\misf=(1-\delta)^{-1}$. This follows from $\sum_{i=1}^\infty \sf_i/\sf_1=1/\sf_1$.

\item $g_n(t)$ in \eqref{gn} is the ccdf of $\sf_n/(1-\sf_{\Sigma}^{(n-1)})$ for $t\geq 1/2$. This is the distribution of $\sf_n$ if
base stations $1$ to $n-1$ did not exist or, equivalently, if the signals from these base stations were 
decoded and cancelled through successive interference cancellation \cite{net:Zhang14tit}.

Some special cases lead to very simple results. For example, setting $t=\delta=1/2$, we have
$2/\pi$, $1/\pi$, $4/(3\pi^2)$, and $1/(2\pi^2)$ for $n=1,2,3,4$, respectively. For $n=1$, in general,
$\bar F_{\sf_{1}}(t)=\sinc(\delta) (t^{-1}-1)^\delta$, $t\geq 1/2$.

\item We can obtain an upper bound on $\E(\sf_1)$ from the fact that for $t<1/2$, \eqref{sic} is an upper bound \cite{net:Zhang14tit}:
\begin{equation}
\E(\sf_1)< \int_0^1 \min(1,g_1(t))\dd t
\label{sf_upper}
\end{equation}
This bound, together with a simulation result, is shown in \figref{fig:mean_sf}. It is apparent that the bound is rather tight.
\item The asymptotic behavior of the cdf of $\sf_1$ as $t\to 0$ is $F_{\sf_1}(t)\sim e^{s^\star(t^{-1}-1)}$ \cite[Thm.~1]{net:Ganti16icc}.
Here $s^\star$ is given by $\,_1F_1(-\delta,1-\delta,-s^\star)=s^{\star\delta}\Gamma(-\delta,s^\star)=0$, where $\Gamma$
is the lower incomplete gamma function. This indicates that the cdf is maximally flat at $t=0$, \ie, all derivatives are $0$.
\end{itemize}

\begin{figure}
\centerline{\epsfig{file=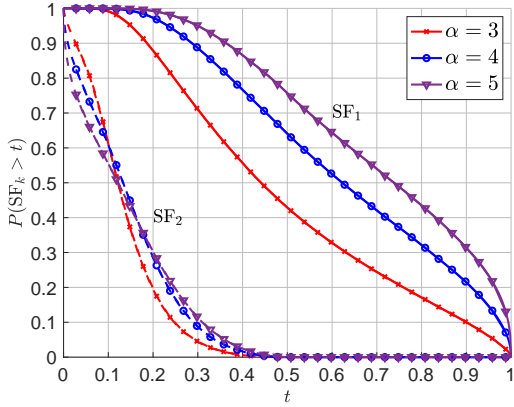,width=.95\figwidth}}
\caption{Ccdfs of $\sf_1$ and $\sf_2$ for $\alpha=3,4,5$. For $t<1/2$, they are obtained by simulation.
For $t\geq 1/2$, the ccdfs of $\sf_1$ are given by $g_1$ in \eqref{gn} while those of $\sf_2$ are $0$.}
\label{fig:sf_no_fading}
\end{figure}

\figref{fig:sf_no_fading} shows the ccdfs of $\sf_1$ and $\sf_2$ for different $\alpha$, partially simulated and partially (for $t\geq 1/2$)
given in \eqref{gn}.
The support of $\sf_n$ is $[0,1/n]$ since the $n$-th largest element cannot exceed $1/n$.
Remarkably, the ccdfs of $\sf_2$ are insensitive to $\alpha$. At $t=1/8$, they are all about $1/2$.
This is explained by the fact that the more dominant $\sf_1$, the smaller $\sf_2$. Thus the gap
between the two ccdfs widens as $\alpha$ increases.

From \cite[Thm.~2]{net:Ganti16twc}, the moments of $\sf_1^{-1}=1+\isr$ are known;
the mean is the MISF,
and $\var(\sf^{-2})=\delta(2-\delta)^{-1}(1-\delta)^{-2}$. Equipped with the moments,
we can derive Markov bounds, such as the lower bound
\[ \bar F_{\sf}(t)\geq 1-\frac\delta{2-\delta} \frac{t^2}{(1-\delta-t)^2},\quad t<1-\delta .\]
However, these are not particularly tight when applied to the SF.

\subsection{Random Base Station Association}
\label{sec:random}
Here we consider the random base station association scheme (RBA)
where, given $\Psi$, the probability of being served by base station $k$ is $\sf_k$.
The ccdf of the resulting SF, denoted by $\widehat{\sf}$, is
\[ \bar F_{\widehat{\sf}}(t)=\E\sum_{k=1}^\infty \sf_{k} \,\one(\sf_{k}>t). \]
It is shown in \cite{net:Keeler14wcl} that $\widehat{\sf}$ has the (standard) beta pdf 
\begin{equation}
   f_{\widehat{\sf}}(t)=\frac{\sin(\pi\delta)}{\pi t^\delta (1-t)^{1-\delta}} 
   \label{random}
 \end{equation}
with mean $1-\delta$, which corresponds to the lower bound in \figref{fig:mean_sf}.
For $\delta=1/2$, this is the arcsine distribution with cdf $F_{\widehat\sf}(t)=2\arcsin\sqrt t/\pi$,
which has the same $\Theta(\sqrt t)$ scaling as the cdf for nearest-neighbor association with Nakagami-$1/2$ fading
(see Subsec.~\ref{sec:fading}). It turns out, surprisingly, that the entire distributions appear to match,
which leads to the following conjecture.

\begin{conjecture}
The SF distribution for nearest-base station association with Nakagami-$1/2$ fading (NBA-$\frac12$) and $\alpha=4$ is
$F_{\sf}(t)=2\arcsin\sqrt{t}/\pi$, $0\leq t\leq 1$.
\end{conjecture}
The evidence supporting the conjecture is that the first $10$ moments of \eqref{random}
and the empirical moments taken over $2\cdot 10^7$ realizations differ by less than $0.03\%$,
and the maximum vertical difference between the arcsine cdf and the empirical one is less than $1/3000$. 

If the conjecture holds, the mean SF for NBA-$m$ increases from $1-\delta$ for $m=1/2$ to
the ``no fading" curve in \figref{fig:mean_sf} as $m\to\infty$.

Comparing the cases of RBA, NBA-$1$, and ISBA, we find:
\begin{itemize}
\item For RBA: $f_{\sf}(0)=\infty$
\item For NBA-$1$: $f_{\sf}(0)=\misr$
\item For ISBA: $f_{\sf}(0)=0$ (and all derivatives are $0$ as well)
\end{itemize}

The ISBA behavior is consistent with the fact that for NBA-$m$, $m\in\N$, the first $m-2$ derivatives of the pdf are zero at $t=0$.
For the tail, $f_{\sf}(t)=\delta\sinc(\delta)(1-t)^{\delta-1}$, $t\to 1$, in all cases.

\section{Conclusions}
The SIR analysis and/or visualization via signal fractions offers several important advantages:
\begin{itemize}
\item Plotting SF distributions (or, equivalently, plotting SIR distributions in MH units) gives the complete
information, no truncation is needed. The asymptotics at low and high SIR are directly visible, and
$\bar F_{\sf}(t)$ near $t=0$ reveals the reliability-rate tradeoff.
\item Due to the bounded support of the SF, all integrals (such as the moments) are guaranteed to be finite.
\item (Generalized) beta approximations are applicable and may lead to new insights.
\item The denominator corresponds to the received (total) signal strength, often abbreviated as RSS. This is the quantity
easily measured at a receiver and also the quantity relevant in energy harvesting. Further, it does not change when
the desired transmitter (such as the serving base station) changes.
\end{itemize}
Focusing on Poisson cellular networks, we have found the BEST approximation for Rayleigh fading and offered a conjecture
on the SF (and thus SIR) distribution with Nagakami-$1/2$ fading. Both would have been unlikely to be found without the
``detour" of using signal fractions.

Lastly, noise can be included by defining the {\em signal fraction with noise} (SFN) as $\sfn\triangleq S/(S+N+I)$, where $N$ is the noise power.
The mapping from the SINR to the SFN is still given by $T$.

\bibliographystyle{IEEEtr}

 \end{document}